\documentclass[aps,prl,groupaddress,amssymb,twocolumn,reprint]{revtex4-1}
\usepackage[utf8]{inputenc}
\usepackage[breaklinks]{hyperref}
\usepackage{graphicx}

\usepackage{natbib}
\usepackage{outlines}
\usepackage{hyperref}

\usepackage{indentfirst}
\usepackage{amsmath}
\usepackage{xfrac}
\usepackage{subfigure}

\usepackage[usenames,dvipsnames]{pstricks}
\usepackage{epstopdf}
\usepackage{pst-grad} 
\usepackage{pst-plot} 

\newsavebox{\astrutbox}
\sbox{\astrutbox}{\rule[-5pt]{0pt}{20pt}}

\newcommand\etal{\mbox{\textit{et al.}}}

\usepackage{ulem}

\newcommand\DEL[1]{\sout{}}     
\usepackage{graphicx}
\usepackage{color}

\begin{document}

\title{Orientation and dynamics of stiff polymeric nanoparticles}



\author{Christophe Brouzet}
\email[]{brouzet@mech.kth.se}
\affiliation{Linné FLOW Centre, KTH Mechanics, KTH Royal Institute of Technology, Stockholm SE-100 44, Sweden}
\affiliation{Wallenberg Wood Science Centre, KTH Royal Institute of Technology, Stockholm SE-100 44, Sweden}
\author{Nitesh Mittal}
\affiliation{Linné FLOW Centre, KTH Mechanics, KTH Royal Institute of Technology, Stockholm SE-100 44, Sweden}
\affiliation{Wallenberg Wood Science Centre, KTH Royal Institute of Technology, Stockholm SE-100 44, Sweden}
\author{L. Daniel Söderberg}
\affiliation{Linné FLOW Centre, KTH Mechanics, KTH Royal Institute of Technology, Stockholm SE-100 44, Sweden}
\affiliation{Wallenberg Wood Science Centre, KTH Royal Institute of Technology, Stockholm SE-100 44, Sweden}
\author{Fredrik Lundell}
\affiliation{Linné FLOW Centre, KTH Mechanics, KTH Royal Institute of Technology, Stockholm SE-100 44, Sweden}
\affiliation{Wallenberg Wood Science Centre, KTH Royal Institute of Technology, Stockholm SE-100 44, Sweden}

\date{\today}

\begin{abstract}

Successful assembly of suspended nanoscale rod-like particles depends on fundamental phenomena controlling rotational and translational diffusion. Despite the significant developments in fluidic fabrication of nanostructured materials, the ability to quantify the dynamics in processing systems remains challenging. Here we demonstrate an experimental method for characterization of the orientation dynamics of nanorod suspensions in assembly flows using birefringence relaxation. The methodology is illustrated using nanocelluloses (cellulose nanocrystals and nanofibrils) as model systems, where the coupling of rotational diffusion coefficients to particle size distributions as well as flow-induced orientation mechanisms are elucidated. Our observations advance the knowledge on key fundamental nanoscale mechanisms governing the dynamics of nanotubes and nanorods allowing bottom-up assembly into hierarchical superstructures.

\end{abstract}


\maketitle

Directed nanoparticle self-assembly is paramount to fabrication of novel materials~\cite{Davisetal2009}. While microfluidics has emerged as a promising tool to accomplish this "bottom-up" approach~\cite{Nunesetal2013,Houetal2017}, it demands scientific understandings concerning nanoparticle dynamics in flow systems. Particularly, controlled assembly of elongated nanoparticles (nanotubes, protein- and polymer-based nanofibrils) into high-performance structural components such as fibres or filaments have recently gained much attention~\cite{Lietal2010,Kiriyaetal2012,NatureCom2014,Kamadaetal2017}. The mechanical performance of these macroscopic material components is given by their nanostructure, predominantly the orientation of the nanoparticles~\cite{Zhouetal2009,Youngetal2010,Kiriyaetal2012,Shinetal2012,Behabtuetal2013,NatureCom2014,Reiseretal2017}. Hydrodynamics can cause local nanoparticle alignment~\cite{Jeffery1922,Harasimetal2013} through shear and extensional flows but our knowledge and models for inter-particle interactions are not sufficient to readily describe the behavior of nanoparticles in flowing suspensions~\cite{FolgarTucker1983,ShaqfehKoch1990,Krochaketal2008,Broederszetal2008,Celzardetal2009}.

Nanoscale assembly processes in flow systems are typically governed by Brownian diffusion, competing with hydrodynamic alignment mechanisms. De-alignment of elongated nanoparticles, here referred to as rotational diffusion, is detrimental to the performance of the macroscopic materials by altering the internal morphology~\cite{Mulleretal2000,Rothetal2003}. Extensive studies on rotational diffusion of monodisperse systems have been carried out~\cite{DoiEdwards1986}, and extended to the polydisperse case~\cite{MarrucciGrizzuti1983,MarrucciGrizzuti1984}, which is typical for biological and chemical systems based on self-assembly of nanosized building blocks~\cite{NatureCom2014,Kamadaetal2017}. However, quantification of rotational Brownian motion is not an easy task and fundamental concepts about motions of anisotropic macromolecules remain largely unexplored~\cite{Hanetal2006,Fakhrietal2010}. Therefore, in-depth characterization of nanoparticle polydispersity and a thorough understanding of the physics behind rotational diffusion will pave the way for flow-based fabrication processes. Existing characterization techniques such as dynamic light scattering~\cite{Levinetal2017}, rheology measurements, electron microscopy, X-ray scattering~\cite{Tanakaetal2017}, orientation relaxation methods~\cite{Rogersetal2005,Rogersetal2005b}, are predominantly limited to observing static systems or require significant investments in instrumentation. A straightforward dynamic characterization technique during fabrication of nanostructured materials will contribute critically to the development of knowledge and technology for controlled assembly in colloidal systems.

Herein, we describe a new experimental methodology that allows real-time assessment of orientation dynamics of nanorods under dynamic flow conditions. This approach is based on a flow-stop technique~\cite{RotDiffTomas}  coupled with thorough analysis of the orientation relaxation~\cite{Rogersetal2005}. Once the flow has reached a steady state, it is rapidly stopped and the relaxation of the particle orientation towards isotropy, depending on the diffusion of the system constituents, is then measured at different locations in the channel. Combining these measurements with a model of rotational diffusion~\cite{DoiEdwards1986,MarrucciGrizzuti1983,MarrucciGrizzuti1984}, this methodology provides information about the typical size of the aligned particles before the flow is stopped. Therefore, the evolution of the length distribution of the aligned particles in the flow can readily be quantified. For the demonstration, we have utilized two types of nanocelluloses as model systems: cellulose nanocrystals (CNC), being almost monodisperse in length, and cellulose nanofibrils (CNF), with a broad length distribution (polydisperse). 

\begin{figure}[t!]
\begin{center}
  \includegraphics[width=0.97\linewidth,clip=]{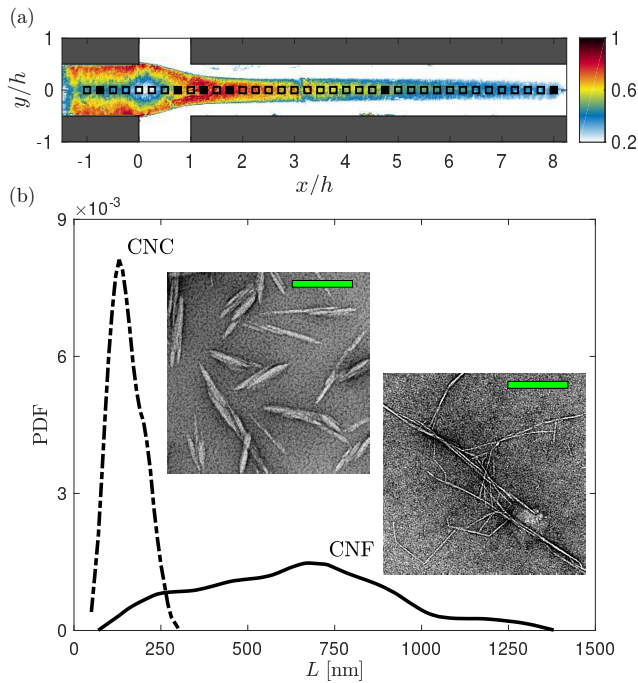}
   \caption{(a): Typical birefringence signal for the CNF suspension. The signal is normalized by its maximum value in the channel. The squares represent the different areas where the birefringence is averaged in space. (b): Length distributions of the CNC (dashed dotted line) and CNF (solid line) suspensions. The insets show TEM images with the scale bars representing $200$~nm.}\label{fig:setup}
\end{center}
\end{figure}

The experimental set-up used in this study consists of a flow-focusing channel~\cite{Nunesetal2013}, each branch having a square cross-section with $h=1$~mm sides as shown in Fig.~\ref{fig:setup}(a) (see Supplemental Material~\cite{Schematic} for a complete description). The nanoparticle suspension flows from the left branch to the right one while two distilled water sheath flows enter in the channel through the top and bottom branches. They focus the nanoparticle suspension into a thread and give rise to an extensional flow, aligning the nanoparticles along the $x$-direction~\cite{Kiriyaetal2012,NatureCom2014}. Fabrication of hierarchical superstructures by self-assembly through the ionic cross-linking of the nanoparticles or by tuning the nanoparticles concentration in the suspensions is conceivable with a similar geometry~\cite{Rammenseeetal2008,Kangetal2012,NatureCom2014,Haynletal2016,Kamadaetal2017,Jahnetal2004,Schabasetal2008}.

The orientation of the nanorods in the flow is visualised through birefringence using polarised optical microscopy~\cite{RotDiffTomas}. The birefringence $B$ gives a quantitative measure of the average orientation~\cite{Schematic}: it is zero for isotropic suspensions and positive for anisotropic ones. A typical CNF birefringence signal is shown in Fig.~\ref{fig:setup}(a). Before the focusing point ($x/h<0$), the maximum birefringence is located on the walls, due to alignement of the nanoparticles with shear~\cite{Jeffery1922}. In the focusing region ($0<x/h<2$), the particles are aligned by the extensional flow and the birefringence reaches its maximum value at around $x/h=1.25$. After the focusing region ($x/h>2$), the thread attains its final shape with no further mechanisms causing alignment. Thus, the particles are relaxing towards isotropy due to rotary diffusion, while being advected by the flow.

The flow is stopped using slider valves~\cite{RotDiffTomas,Schematic}. For both the suspensions (CNC and CNF), the birefringence decay $B(t)$ is observed at different locations along the centreline of the channel, marked by the black squares in Fig.~\ref{fig:setup}(a). The camera is used to record $2000$~images at $5000$~fps for the CNC and $30000$~images at $1000$~fps for the CNF. Thus, the decay dynamics is sampled on more than $3$~decades for CNC and $4$~decades for CNF. 

The sample preparation is described in Supplemental Material~\cite{SamplePreparation}. The length and diameter of $200$ nanoparticles were measured using transmission electron microscopy (TEM) and atomic force microscopy (AFM), respectively. The length distributions of the corresponding samples are plotted in Fig.~\ref{fig:setup}(b), together with representative TEM images. The CNC (dashed dotted line) are monodisperse with a mean length of $165$~nm while the CNF (solid line) are clearly polydisperse with a most probable length of $670$~nm. The diameter is almost constant in each sample and around $15$~nm for the CNC and $3$~nm for the CNF (see data in the Supplemental Material~\cite{WidthData}). The dry weight concentrations of the suspensions used in this study are $41$~g/l for CNC and $3$~g/l for CNF. Nanocelluloses have been chosen as model systems as they can be treated as rigid rods due to an elastic modulus of $130-140$~GPa~\cite{Klemmetal2011}. The estimated persistence lengths of the samples~\cite{Fakhrietal2010}, $150~\mu$m for the CNF and even higher for the CNC, are much larger than the length of the particles, which ranges from a few hundred nanometers to $1~\mu$m. 

\begin{figure}[t!]
\begin{center}
  \includegraphics[width=0.9\linewidth,clip=]{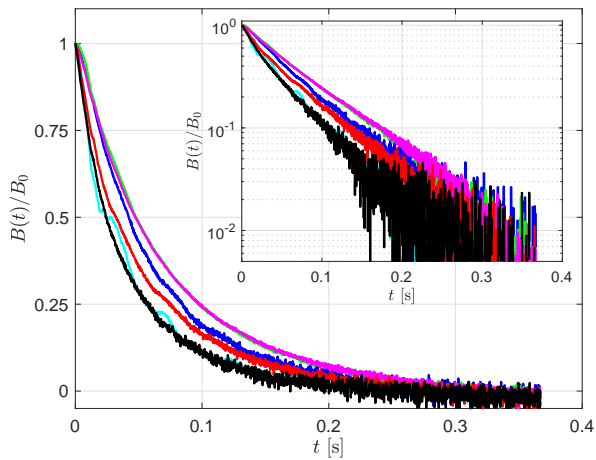}
   \caption{Normalized birefringence decays for the CNC suspension at different positions in the channel: $x/h=-1$ (blue), $0.75$ (light blue), $1.25$ (green), $1.75$ (magenta), $4.75$ (red) and $8$ (black). The inset shows the same plot with a logarithmic scale for the vertical coordinates.}\label{fig:CNC_decays}
\end{center}
\end{figure}

To correlate the return-to-isotropy of the samples with a given particle length distribution, it is necessary to model the effect of Brownian motion. Doi and Edwards~\cite{DoiEdwards1986} developed a model for monodisperse rods which depends on the length $L$, diameter $d$ and concentration $c$ of the particles. For $cL^3 \ll 1$ (dilute regime), the rods are free to rotate without any inter-particle interactions, while for $1 \ll cL^3 \ll L/d$ (semi-dilute regime), the effects of particle interactions on the particle dynamics become significant due to volume exclusion~\cite{DoiEdwards1986} and network formation~\cite{Broederszetal2008,Celzardetal2009}. With a dry weight concentration of $41$~g/l ($cL^3 \approx 4.2$), the CNC suspension belongs to the semi-dilute regime. The rotary diffusion coefficient of a rod in an isotropic semi-dilute system is
\begin{equation}
D_r=\frac{\beta}{(cL^3)^2} \frac{k_B T \ln(L/d)}{3 \pi \eta L^3},\label{eq:monodisperse}
\end{equation}
where $T$~is the temperature, $k_B$~the Boltzmann constant, $\eta$~the solvent viscosity and $\beta$~a numerical factor. The birefringence relaxation towards isotropy is predicted to be exponential with a typical time scale of $\tau=1/(6 D_r)$~\cite{DoiEdwards1986}. Figure~\ref{fig:CNC_decays} shows the birefringence relaxation of the CNC suspension at different locations in the channel, both in a lin-lin and log-lin plot (inset). All the curves are normalized by their initial value, $B_0$. They almost collapse and the inset indicates an exponential decay, as expected. This illustrates that independent of the initial alignment, only one time scale can be associated to the relaxation of the monodisperse CNC particle system, in agreement with Doi and Edwards theory~\cite{DoiEdwards1986}.

\begin{figure}[t!]
\begin{center}
  \includegraphics[width=0.95\linewidth,clip=]{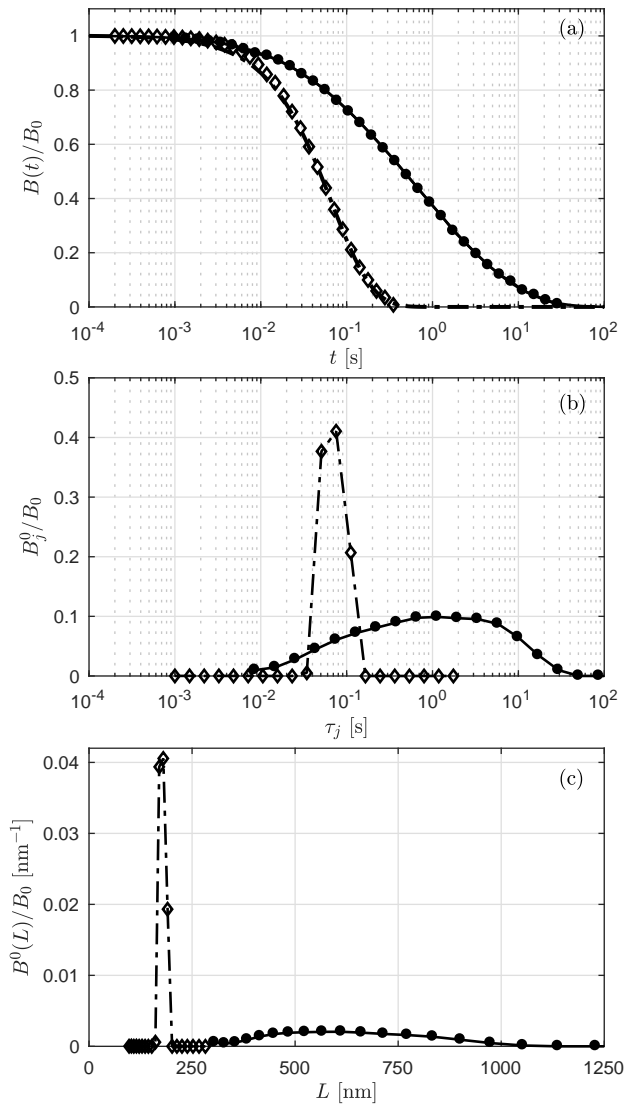}
   \caption{(a): Normalized birefringence decays for CNC (empty diamonds) and CNF (filled circles), at $z/h=1.75$. The fit of each curve using the inverse Laplace transform method is shown as dashed dotted line for CNC and solid line for CNF. (b) (respectively (c)): Normalized contributions $B_j^0/B_0$ (resp. $B^0(L)/B_0$) to the different time scales $\tau_j$ (resp. length scales $L$) in the decay signals in panel (a) for CNC (dashed dotted line and empty diamonds) and CNF (solid line and filled circles).}\label{fig:method}
\end{center}
\end{figure}

The diffusion model~(\ref{eq:monodisperse}), valid for the monodisperse CNC suspension, needs to be extended in order to describe the dynamics of the polydisperse CNF suspension, which is more realistic for general nanoparticle systems. A model for polydisperse systems has been proposed by Marrucci and Grizzuti~\cite{MarrucciGrizzuti1983,MarrucciGrizzuti1984}. In such systems, the concentration distribution $\tilde c$ depends on the rod length $L$ and the transition between dilute and semi-dilute regimes is defined by the entanglement length 
\begin{equation}
L_*=\left(\int_0^{+\infty} \tilde c(L) L~\textrm{d}L \right)^{-1/2}.
\end{equation}
Rods shorter than $L_*$ are considered to be in the dilute regime while rods longer than $L_*$ are in the semi-dilute regime. For the CNF suspension, $L_* \approx 60$~nm, reflecting that all CNF are in the semi-dilute regime (see Fig.~\ref{fig:setup}(b)). In this regime, the diffusion coefficient of a rod of length $L$ depends on its interactions with other rods of different lengths, which may have different orientation distributions~\cite{MarrucciGrizzuti1983,MarrucciGrizzuti1984}. Therefore, the relaxation dynamic of one rod is coupled to the others. For a polydisperse system close to isotropy, the diffusion coefficient for a rod of length $L$ can be written as~\cite{MarrucciGrizzuti1984,ApproximationDerivation}
\begin{equation}
D_r(L)=\frac{\beta k_B TL_*^4}{\eta L^7}\Gamma(L),\label{eq:polydisperse}
\end{equation}
where
\begin{equation}
\Gamma(L)=\frac{\int_0^{+\infty} \tilde c(L'){L'}\textrm{d}L' }{\left(\int_0^{L} \tilde c(L'){L'} \left(\frac{L'}{L}\right)^3\textrm{d}L' + \int_L^{+\infty} \tilde c(L'){L'}\textrm{d}L' \right)}.\label{eq:gamma}
\end{equation}
Equation~(\ref{eq:polydisperse}) is similar to Eq.~(\ref{eq:monodisperse}), with the concentration dependency hidden into the entanglement length $L_*$ and with the logarithm of the aspect ratio replaced by the factor $\Gamma(L)$; the latter is the correction factor due to interactions between rods of different lengths. For the smallest rods, $\Gamma$ is strictly equal to $1$ while it remains of order $1$ for longer rods (see details in the Supplemental Material~\cite{ApproximationDerivation}). Using this isotropic approach, the dynamics of each rod is now decoupled and takes place at its own time scale $\tau_L=1/(6 D_r(L))$, depending on its length $L$ and the concentration distribution $\tilde c$. As the total birefringence is the sum of the contributions of each length component~\cite{Rogersetal2005,Rogersetal2005b}, the birefringence relaxation signal is therefore given by 
\begin{equation}
B(t)=\int_0^{+\infty} B^0(L) \exp\left(-6 D_r(L) t\right) \textrm{d}L.\label{eq:decay}
\end{equation}
Here $B^0(L)$ is the contribution of the nanoparticles of length $L$ to the total birefringence signal $B_0$ before the flow is stopped. This leads to a relaxation towards isotropy with multiple time scales~\cite{Rogersetal2005,Rogersetal2005b,RotDiffTomas,Chowetal1985}, visible for the CNF in Fig.~\ref{fig:method}(a) (filled circles). The difference with the exponential relaxation of the CNC, also plotted in Fig.~\ref{fig:method}(a) (empty diamonds), is significant and highlights the strong coupling between the orientation relaxation of a system and its length distribution.

\begin{figure*}
\begin{center}
  \includegraphics[width=0.95\linewidth,clip=]{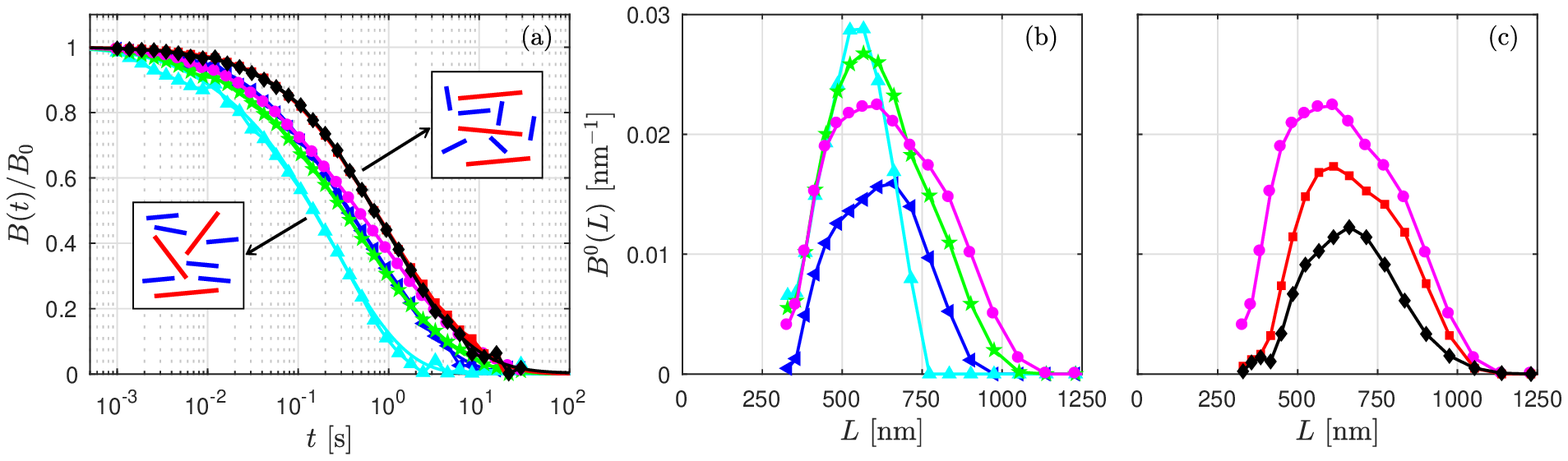}
   \caption{(a): Normalized birefringence decays for CNF. The symbols are experimental data and the solid lines are the inverse Laplace transform fit. The colors are for different positions in the channel: $x/h=-1$ (blue), $0.75$ (light blue), $1.25$ (green), $1.75$ (magenta), $4.75$ (red) and $8$ (black). The two insets are schematics of the orientation of the short (blue) and long (red) fibrils \emph{before} the stop, in the focusing region and far downstream. (b) and (c): Contributions $B^0(L)$ to the different length scales $L$ upstream ($x/h\leq1.75$) and downstream ($x/h \geq 1.75$), respectively. The colors used are the same as in panel (a). Error bars~\cite{ErrorBars} are approximately of the same size of the symbols used in panels (b) and (c).}\label{fig:CNF_decays}
\end{center}
\end{figure*}

This coupling allows us to extract information on the length distribution. Indeed, Eq.~(\ref{eq:decay}) is a Laplace transform of $B^0(L)$ and this quantity can be estimated by inverting $B(t)$, using a method described by Rogers \etal~\cite{Rogersetal2005,Rogersetal2005b}. The method is illustrated in Fig.~\ref{fig:method} with two examples, one for CNC (empty diamonds) and one for CNF (filled circles). Figure~\ref{fig:method}(a) shows the birefringence $B/B_0$ as a function of time in a lin-log plot, where the signals have been resampled on $100$~log-spaced points. The log-scale is necessary because a small change in length is amplified by the power $7$ for time, as shown in Eq.~(\ref{eq:polydisperse}). To perform the inverse Laplace transform, a set of $20$~log-spaced time scales $\{\tau_j\}$ is chosen within the range $[5 t_\textrm{min}; 5 t_\textrm{max}]$, with $t_\textrm{min}$ the time interval between two images and $t_\textrm{max}$ the duration of the full acquisition. By inverting the signals in Fig.~\ref{fig:method}(a), we obtain a set of $20$~$B_j^0$ associated to the $\{\tau_j\}$. This set $\{B_j^0\}$ is a discrete version of $B^0(L)$. When $B_j^0 \neq 0$, the associated time scale $\tau_j$ contributes to the decay. For the CNC signal, Fig.~\ref{fig:method}(b) exhibits a sharp peak while the broad distribution for the CNF signal spreads over more than $3$~decades in time. This highlights that many time scales contribute to the CNF decay, as can be expected due to the polydispersity. The time scales $\{\tau_j\}$ are then converted to length scales $L$ using Eq.~(\ref{eq:polydisperse}) and $\tau(L)=1/(6 D_r(L))$ while the discrete set $\{B_j^0\}$ is turned into a continuous distribution $B^0(L)$. Note that this conversion is made by neglecting the factor $\Gamma$ (of order $1$) with respect to the dominant contribution given by the rod length dependency (to the power $7$) in Eq.~(\ref{eq:polydisperse}). The numerical factor $\beta$ given in Eq.~(\ref{eq:polydisperse}) has also to be adapted. For the CNF, it has been set to $\beta=10^3$, which is the order of magnitude reported by other experiments~\cite{DoiEdwards1986}. For the CNC, $\beta=0.5$ matches well with the length distribution. The discrepancy between the two values reflects that CNC and CNF suspensions form three-dimensional networks differently due to their length distributions and morphology. Nevertheless, we finally obtain in Fig.~\ref{fig:method}(c) an estimation of the different lengths contributing to the birefringence, i.e. the lengths aligned by the flow before it was stopped. Note that the horizontal axis in Fig.~\ref{fig:method}(c) is linear and within the range of the length distributions given in Fig.~\ref{fig:setup}(b).

By using systematically the method presented in Fig.~\ref{fig:method} at different channel locations, it is now possible to quantify the evolution of the length distribution of the \emph{aligned fibrils} along the channel and therefore to assess the orientation dynamics of the suspension in the flow. Figure~\ref{fig:CNF_decays}(a) shows normalized birefringence decays for CNF at different positions along the channel, revealing different time scales in the dynamics. The contributions $B^0(L)$ obtained after the inverse Laplace transform of these decays are presented in Figs.~\ref{fig:CNF_decays}(b) and~(c) for $x/h\leq1.75$, where fibrils are aligned by the extensional flow, and for $x/h\geq1.75$, where the alignment mechanism vanishes. The magenta curve ($x/h=1.75$) is repeated in both figures for an easier comparison. 

The distribution of contributions $B^0(L)$ to the birefringence before the focusing is represented with blue triangles in Fig.~\ref{fig:CNF_decays}(b). It is modified greatly after the focusing. First, it becomes narrower and the maximum shifts towards shorter particles (light blue triangles). Subsequently, longer particles contribute more (green stars and magenta dots), indicated by the maximum and the right flank of the distribution shifting towards longer particles. This means that, during the alignment process, short fibrils are aligned first followed by alignment of longer particles. Further downstream, where no alignment mechanisms are present (Fig.~\ref{fig:CNF_decays}(c)), rotary diffusion reduces the contributions at all lengths, especially for short particles. 

These findings allow us to establish a comprehensive understanding of the orientation dynamics in microfluidic channels in general, and specifically in the extensional flows. Short particles are quickly oriented by the flow while more time is needed to align long ones. Furthermore, short particles de-align faster and their motion is therefore difficult to control during the assembly. This scenario can be also deduced from Fig.~\ref{fig:CNF_decays}(a), used as a diagnostic plot: aligned short fibrils in the focusing region introduce short time scales in the decay (light blue triangles) while aligned long fibrils further downstream exhibit longer time scales (red squares and black diamonds). These findings may have strong implications on the performance of the macroscopic nanocellulose structures fabricated via flow-based assembly~\cite{NatureCom2014,Waltheretal2011,Lundahletal2016}. The typical time for the transition to colloidal glassy-state achieved with the flow focusing geometry of similar dimensions is estimated to be around $3.3$~s~\cite{NatureCom2014}, i.e. longer than the time found here for the de-alignement of the shortest fibrils (see Fig.~\ref{fig:CNF_decays}(a)). Therefore, the hierarchical structures manufactured using similar approaches may be composed of long aligned particles embedded into an isotropic matrix of shorter particles. 

The orientation dynamics therefore exhibits a strong dependency on the length of the fibrils. This can be explained by the entanglement of the CNF which increases with their length, in the formed three-dimensional network. Indeed, during hydrodynamic alignment, all fibrils tend to rotate but the concentration required for the fabrication of hierarchical structures is typically sufficiently high for the fibrils to quickly collide and for their motion to be hindered. Such collisions tend to re-orient the fibrils and slow down the alignment process~\cite{FolgarTucker1983}. Thus, at a given total concentration small fibrils endure a less hindered motion, experience fewer collisions and a quicker alignment than the long ones, despite higher diffusion coefficients. When the alignment mechanisms are no longer effective, the differences in diffusion coefficients, due also to the entanglement, result in faster de-alignment of the short fibrils. It should be noted that contributions from the minimal and maximal lengths of the fibrils in the polydisperse sample (see Fig.~\ref{fig:setup}(b)) could not been accounted for in Fig.~\ref{fig:CNF_decays}(b) and (c). This highlights the fact that these fibrils are either too short, having a diffusion time scale smaller than the alignment time scale and therefore an inefficient alignment, or too long causing a highly entangled network hindering alignment.

To summarize, combining flow-stop experiments and inverse Laplace transforms, our observations on the Brownian effects of nanorod suspensions provide exquisitely detailed information about the diffusive properties of anisotropic nanoobjects and the subtle interplay between alignment and diffusion mechanisms. Besides characterizing the nanoparticle dynamics under dynamic flow conditions, this technique is able to reveal that the orientation dynamics is strongly dependent on the nanoparticle length distribution and highlights 
the importance of the network formation due to the inter-particle interactions, which is challenging to understand theoretically. To the best of our knowledge, this is the first of its kind methodology that allows such assessment and characterization of dynamic colloidal systems.

\begin{acknowledgments}

The Wallenberg Wood Science Centre at KTH is acknowledged for providing the financial assistance. The authors are thankful to Dr. A. Boujemaoui for providing the CNC suspension and to Prof. H. A. Stone for valuable comments. Dr. T. Rosén is acknowledged for experimental assistance and helpful discussions.

\end{acknowledgments}

\end{document}